\documentclass{aa}
\usepackage{graphicx}
\begin{document}
   \title{RFI excision using a higher order statistics analysis of the power spectrum}

   \author{P. A. Fridman}

   \offprints{P. Fridman, \email{fridman@nfra.nl}}

      \institute{ASTRON, Postbus2 , 7990AA, Dwingeloo, The Netherlands\\
	\email{fridman@nfra.nl}
             }
   \date{Received ; accepted }

   \abstract{
   A method of radio frequency interference (RFI) suppression in radio astronomy spectral observations is described based on the analysis of the probability distribution of an instantaneous spectrum. This method allows the separation of the gaussian component due to the natural radio source and the non-gaussian RFI signal. Examples are presented in the form of 
   \keywords{Power spectrum -- probability distribution --
                RFI excision}        
   }

   \maketitle
%

\section{ Introduction}

The increasing radio frequency interference (RFI) from transmissions and spurious signals of commercial telecommunication applications limit the real sensitivity of radio telescopes at certain frequencies. This radio-ecological environment at radio observatories appears to be worsening each year, while the observing systems get more sensitive. Further improvements in antenna quality and in receiver parameters do not give the expected results because of limitations due to RFI. Therefore, it is very difficult to reach the design sensitivity, especially at low frequencies, even for radio telescopes situated in secluded areas (Spoelstra\cite{spoelstra97}; Kahlmann  \cite{kahlmann99}; Cohen \cite{cohen99}).

 Several methods of RFI excision have been proposed during recent years. The successful application of these methods depends on the type of radio telescope, type of observations, type of RFI, and there is no universal RFI mitigation algorithm.

1. {\bf Type of radio telescope} - single dish, connected radio
intererometer, VLBI. Single dish radio telescopes are especially
vulnerable to RFI. Radio interferometric systems suffer from RFI in a less
degree, (Thompson \cite{thompson82}), but auxiliary noise power measurements
made at each antenna for calibration purposes are  distorted by RFI, as for the single dish case.

2. {\bf Mode of observations} - continuum or spectral. In continuum
observations it is possible to sacrifice some part (5-10\%) of the data stream
polluted by a strong RFI and save the remained part without significant loss of
sensitivity. In spectral observations RFI and signal-of-interest may occupy
the same spectral region, so it is impossible to delete this
particular part of the spectrum.

3. {\bf Type of RFI} - impulse-like bursts, narrow-band \ or wide-band.
Theoretical consideration and experimental data show that RFI may be basically represented as the superposition of two types of waveforms: {\it impulse-like bursts and long narrowband} and {\it strongly correlated random oscillations} (Middleton \cite{middleton72,middleton77}; Lemmon \cite{lemmon97}). RFI of the first type can be effectively suppressed using fast real time thresholding before averaging at the total power detector or correlator
(Fridman \cite{fridman96}; Weber et al. \cite{weber97}). This is the processing in {\it temporal domain}.

RFI of the second type should be processed in the {\it frequency domain}, where there is a good contrast between a quasi-continuum spectrum of the {\it system noise plus radio source} signal and the {\it narrow-band RFI} signal. RFI in continuum observations may be excised using adaptive filtering methods (Fridman \cite{fridman98}), such that the parts of the spectrum affected by RFI are used with a smaller weighting than the other ''unpolluted'' spectral regions during the averaging in the time-frequency plane. This sort of processing cannot be applied in spectral line observations, where RFI may contaminate the same spectral region that is of interest for radio astronomer. In (Ellington, Bunton and Bell \cite{ellington00}) RFI from navigational satellites GLONASS was parametrically modelled and the model of RFI signal was subtracted from the radio telescope output so that the 1612MHz OH spectral line was not distorted.

Adaptive interference cancelation (similar to
adaptive noise cancellation, Widrow \&Stearns \cite{widrow85}) can be useful for
single dish observations with an auxuliary reference antenna,
pointed "off-source". The use of an auxiliary reference channel provides some benefits (Barnbaum \& Bradley\cite{barnbaum98}), but not all radio telescopes are or can be equipped with such a reference channel. Furthermore, the effectiveness of this adaptive noise canceling strongly depends on the sensitivity of the reference channel. The reference channel should at least be equally sensitive than the main channel for successful RFI suppression (Fridman \cite{fridman97,fridman00}).

A new method of post-correlation RFI removing was proposed in (Briggs, Bell and Kesteven \cite{briggs00}). Cross-correlation products of four signals (two main channels  and two reference channels) are processed, using closure complex amplitude relations. RFI  signal is estimated and subtracted from the outputs of main channels. So RFI suppression can be made {\it off-line} with the averaged data, see also (The Elizabeth and Frederick White Conference on
Radio Frequency Interference Mitigation Strategies \cite{white99}).

Methods of spatial filtering in the temporal and frequency domains applicable for radio interferometers are discussed elsewhere (Leshem \& van der Veen \cite{leshem99}). Spatial filtering methods use difference in
the direction-of-arrival of radio source  and
RFI. The RFI emission of well spatially
localized sources could be suppressed in multielement radio interferometers
using adaptive array philosophy, when zeros of a synthesized antenna pattern
coincide with directions-of-arrival of undesirable signals (adaptive nulling techniques).
This approach is popular while considering new projects of superlarge phased
array radiotelescopes, (van Ardenne \cite{ardenne96}; Bregman \cite{bregman00}). However,
there are some limitations in using this techniques for the existing large radio
interferometers like WSRT, VLA, GMRT: very sparse antenna arrays, only-phase
computer control, complex impact on image postprocessing.
Spatial filtering is effective when RFIs are strongly correlated at a radio telescope antennas (like in a half-wavelength phased array). RFI at the adjacent antennas (hundreds
meters separation) are highly correlated, but when the distance between
antennas is more than 1km (and especially for bases 10-100km) RFI at these
points are less correlated, and effectiveness of the adaptive nulling is not
so high, as for the continuous-like  half-wavelength phased arrays.
RFI at the sites of VLBI system, separated by hundreds and thousends of
kilometers, are practically uncorrelated. So the  impact of  an RFI is an increased variance at
correlator output, which is noticeable in the case
of strong RFI (comparable with the system noise power).

In this paper, I consider a method of RFI excision based on the analysis of the probability distribution of the power spectrum at the radio telescope output. System and radio source noise generally has a gaussian distribution with a zero mean. If a Fourier transformation is applied to such an ''ideal'' signal, the real and imaginary components in every spectral bin are also gaussian random values with zero mean. The instantaneous power spectrum, which is a square of magnitude of the complex spectrum, has an exponential distribution (or, in other terms, a chi-squared distribution with two degrees of freedom). Any changes of the input signal of this model, due to the presence of an RFI, yield a change of its statistics. The power spectrum in this case has a non-central chi-squared distribution with two degrees of freedom. Radio astronomy observations generally employ the averaging of the power spectrum over some integration period, which results in a conversion of the probability distribution of the averaged power spectrum into a gaussian distribution.

Analysis of sample probability distribution {\bf before averaging} allows an effective separation of these signals by means of real-time processing facilities employing digital signal processing. The method may be effective for single dish  observations without reference channel. The description of this separation by using the analysis of the power spectrum distribution will be presented below. I will also present some computer simulation examples, and some results from test observations at the Westerbork Synthesis Radio Telescope (WSRT).
\section{ Power spectrum probability distributions and moments}

A radio telescope receiver output signal is described as follows:
\begin{equation}
  r(t)=x_{sig}(t)+x_{sys}(t)+x_{RFI}(t),
\end{equation}
where $x_{sys}(t)$ is a system noise signal (receiver, feed, antenna), $x_{sig}(t)$ is the radio source noise signal with some spectral features to be estimated, and $x_{RFI}(t)$ is the RFI waveform. $x_{sys}(t)$ and $x_{sig}(t)$ are each supposed to be random processes with gaussian probability distributions, zero mean and variances $\sigma _{sys}^{2}$ and \ $\sigma _{sig}^{2},$ respectively. However, $x_{RFI}(t)$ is a narrow-band interference signal with variance $\sigma _{RFI}^{2}$. After Fourier transformation on a time interval $T$, we get the instantaneous complex spectrum from eq.(1):
\begin{eqnarray}
   S(f)=\int_{0}^{T}r(t)\cos 2\pi ftdt+ j\int_{0}^{T}r(t)\sin 2\pi ftdt=\nonumber\\
 =S\_re(f)+jS\_im(f)
\end{eqnarray}

Probability distributions of the real part of the spectrum $S\_re(f)$ and the $\ $imaginary part $S\_im(f)$ are also gaussian with a mean ${A}(f)$ and a variance $\sigma _{n}^{2}(f)+{A}(f)_{n}^{2},$ where component ${A}(f)$ is determined by the RFI and the $\sigma _{n}^{2}(f)$ is determined by system + signal spectral densities. A power spectrum (periodogram) is written as:
\begin{equation}
P(f)_{T}=S(f)\overline{S(f)},
\end{equation}
which corresponds to an envelope quadratic detector operation, the overbar denotes complex conjugation. Here $P(f)_{T}$ at each frequency bin $f$ has the probability distribution of the square of the envelope of a randomly phased sine wave and a narrow-band gaussian process
({\it noncentral} $\chi ^{2}$ {\it distribution with two degrees of freedom}):
    \begin{equation}
w(P)=\frac{1}{2\sigma _{n}^{2}}e^{-\frac{P+{A}_{RFI}^{2}}{2\sigma _{n}^{2}}}I_{0}(\frac{{A}_{RFI}\sqrt{P}}{\sigma _{n}^{2}}),P>0,
    \end{equation}
which in the absence of RFI is an {\it exponential distribution}:
\begin{equation}
w(P)=\frac{1}{2\sigma _{n}^{2}}e^{-\frac{P}{2\sigma _{n}^{2}}}.
\end{equation}

After averaging M periodograms (3), $P_{aver}(f) = \sum_{m=1}^{M}P_{m}(f)$, we have for $P_{aver}$ the {\it non-central} $\chi ^{2}$ {\it distribution with} $2M$ {\it degrees of freedom}, (Whalen \cite{whalen71}):

\begin{eqnarray}
w(P_{aver})=\frac{1}{2\sigma _{n}^{2}}(\frac{P\sigma _{n}^{2}}{M{A}_{RFI}^{2}})^{\frac{M-1}{2}}\times\nonumber\\
\exp (-\frac{P}{2\sigma _{n}^{2}}-\frac{M{A}_{RFI}^{2}}{2\sigma _{n}^{2}})I_{M-1}(\frac{{A}_{RFI}\sqrt{PM}}{\sigma _{n}^{2}})
\end{eqnarray}

and in the absence of RFI:

\begin{equation}
w_{0}(P_{aver})=\frac{1}{(2\sigma _{n}^{2})^{M}\Gamma (M)}P^{M-1}e^{-P^{2}/2\sigma _{n}^{2}}.
\end{equation}

When the noncentral parameter ${A}_{RFI}/\sigma_{n} >1$, both distributions (4) and (6) tend to acquire a gaussian form. For big $M$ (central limit theorem) the distribution (6) also is transformed into a gaussian. This near-gaussian distribution characterizes the power spectra at the outputs of common spectrum analyzers. Figure 1a shows how dramatically the distribution (4) is transformed, while ${A}_{RFI}/\sigma_{n}$ grows from zero to 5. Figure 1b represents this kind of transformation with periodogram averaging applied. Figure 2a shows several sections of the three-dimensional presentation of Fig. 1a.

\begin{figure}
\includegraphics[width=8cm]{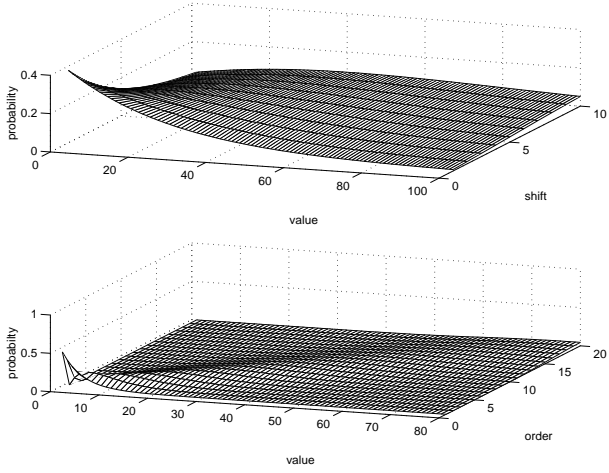}
\caption{The non-central $\chi ^{2}$ probability distribution as a
function of a)(${A}_{RFI}/\sigma_{n} $) and b) the order $M$.}
\end{figure}

One observes that the presence of RFI changes the form of the distribution $w(P)$ of the unaveraged periodogram. This property may be used for separating the (system + signal) noise component $\sigma _{n}^{2}(f)$ from the RFI component ${A}_{RFI}(f)$.

The most promising method is to obtain an estimate of the sample probability distribution $\widehat{w}[P(f)]$ (histogram), and subsequently estimate the $\widehat{\sigma}_{n}^{2}(f)$ and $\widehat{A}_{RFI}(f)$, which best fit this sample distribution (symbol $\widehat{ }$ denotes sample average). However, performing this processing in {\it real-time} with a sufficiently wide band (at the MegaHertz level) is a difficult task even for the most powerful modern Digital Signal Processors (DSPs).

A more attractive practical approach is to use the sample moments of the power spectrum. The first four statistical parameters, {\it mean, variance, skewness and excess}, for distribution (4) are:
\begin{eqnarray}
mean=2\sigma _{n}^{2}+{A}_{RFI}^{2},\\
var=4\sigma _{n}^{4}+4\sigma _{n}^{2}{A}_{RFI}^{2},\\
skew=\frac{2\sigma _{n}^{2}+3{A}_{RFI}^{2}}{[\sigma _{n}^{2}+{A}_{RFI}^{2}]^{3/2}}\sqrt{\sigma _{n}^{2}},\\
\qquad exc=6(\sigma _{n}^{2}+2{A}_{RFI}^{2})\frac{\sigma _{n}^{2}}{(\sigma _{n}^{2}+{A}_{RFI}^{2})^{2}}.
\end{eqnarray}

\begin{figure}
\resizebox{\hsize}{10cm}{\includegraphics{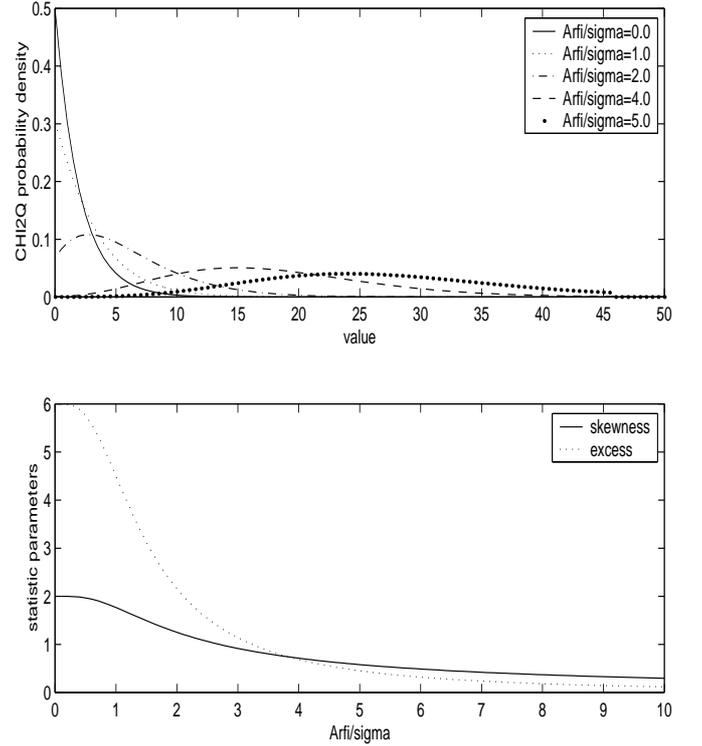}}
\caption{a) The non-central $\chi ^{2}$ probability distribution as a function of a) the value of the noncentral parameter (${A}_{RFI}/\sigma $) and b)the skewness and excess as a functions of ${A}_{RFI}/\sigma_{n}$.}
\end{figure}

In the absence of RFI, the $mean$=$2\sigma _{n}^{2}$, the variance = $4\sigma_{n}^{4}$, the skewness = $2$, and the excess = $6$. For strong RFI, the $skewness \rightarrow 0$ and the excess $\rightarrow 0$, which is a good indicator of the presence of RFI. Fig. 2b shows this dependence of skewness and excess on the parameter ${A}_{RFI}\ /\sigma _{n}$.

Having the sample values of  $\widehat{mean},\widehat{var},\widehat{skew},\widehat{exc}$, this redundant system of nonlinear equations can be solved with respect to $\sigma _{n}^{2}$ and ${A}_{RFI}^{2}$, with the obvios constraints $\sigma _{n}^{2}>0, {A}_{RFI}^{2}>0$. Analysis of the "solution surface" shows that the solution is unique. Even on the basis of only the first two equations, it is easy to derive that at each frequency
\begin{eqnarray}
\widehat{{A}_{RFI}^{2}}=\sqrt{\widehat{mean}^{2}-\widehat{var}},\\
\widehat{\sigma }_{n}^{2}=(\widehat{mean}-\widehat{{A}_{RFI}^{2}})/2.
\end{eqnarray}

The value $2\widehat{\sigma }_{n}^{2}(f)$ as a function of frequency provides "clean" power spectrum.

Sometimes radio astronomers use a cross-correlation spectrum instead of the auto-correlation spectrum, with the aim to suppress large system noise spectral component. The probability distribution and the statistical parameters for this case are given in the Appendix.
\begin{figure}
 

\setlength{\unitlength}{0.1mm}
\fbox{\begin{picture}(850,600)
\put(50,450){\framebox(100,100){Filter}}
\put(150,500){\line(1,0){100}}
\put(250,500){\vector(0,-1){130}}
\put(10,560){\small{system noise}}

\put(5,500){\vector(1,0){45}}
\put(50,270){\framebox(100,100){Filter}}
\put(10,380){\small{signal noise}}

\put(5,320){\vector(1,0){45}}
\put(150,320){\vector(1,0){50}}
\put(200,270){\framebox(100,100){Adder}}
\put(5,150){\line(1,0){245}}
\put(5,160){\small{RFI}}

\put(300,320){\vector(1,0){50}}
\put(500,320){\line(1,0){40}}
\put(540,320){\vector(0,1){130}}
\put(540,320){\vector(0,-1){120}}

\put(250,150){\vector(0,1){120}}
\put(350,270){\framebox(150,100)}
\put(365,340){Fourier}
\put(350,300){transform}
\put(450,450){\framebox(150,100)}
\put(460,525){Moment}
\put(450,480){\small{estimation}}
\put(600,500){\vector(1,0){50}}

\put(650,450){\framebox(150,100)}
\put (670,520){Power}
\put (655,480){spectrum}
\put(725,450){\vector(0,-1){390}}

\put(450,100){\framebox(150,100)}
\put(470,170){Power}
\put(455,130){spectrum}
\put(525,100){\vector(0,-1){40}}
\put(400,75){"dirty"}
\put(400,50){power}
\put(400,25){spectrum}
\put(610,75){"clean"}
\put(610,50){power}
\put(610,25){spectrum}

\end{picture}}

\caption{A block diagram of the computer simulations.}
\end{figure}
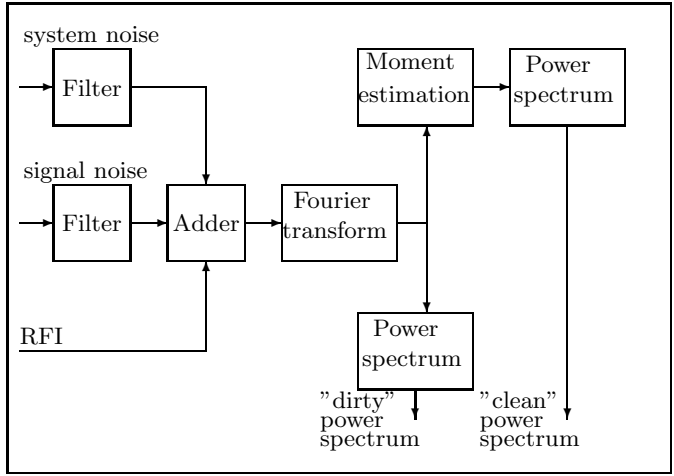

\section{Computer simulations}
Some computer simulations have been done with the aim to investigate how effective the separation of the gaussian noise component and the non-gaussian RFI can be.  Figure 3 represents a block diagram of the simulation program. The three signal components are as follows:  (1) the system noise is a sequence of gaussian random values passed through a passband Butterworth filter with cutoff frequencies [0.1, 0.9] and with order 7, (2) the signal noise imitating a spectral line is also represented as a sequence of gaussian random values passed through a Butterworth filter with cutoff frequencies [0.3 0.35] and with order 5, and (3) the RFI waveforms consists of a superposition of two strong {\it chirp} (frequency modulated sinusoidal wave) signals. One of the RFI signals coincides in frequency with the position of the ''spectral line''. These three signals are added and applied to a spectrum analyzer (FFT block), where an instantaneous complex spectrum is calculated. This operation imitates a discrete digital spectral analysis with L frequency channels. The generation of the mixture {\it noise + RFI} and of the Fourier transform was repeated M times. The upper part of the block diagram after the FFT block corresponds to the proposed RFI excision method, where statistical parameters (mean, variance, skewness, and excess) are calculated at each frequency bin for the M samples. Subsequently the estimates $\widehat{\sigma }_{n}^{2}(f_{l})$, for each frequency channel $l=1...L$ were calculated.

The lower part of the block diagram in Fig. 3 shows the common way of doing spectral analysis: calculating of the power spectrum  $P(f)_{T}=S(f)\overline{S(f)}$ and subsequent averaging. At this point the two versions of the power spectrum $\widehat{\sigma }_{n}^{2}(f_{l})$ are available: one without RFI excision, and another with RFI excision. Figures 4-7 show the results of the some computer simulations. Fig. 4a is an example of an input power spectrum without RFI with its input signal waveform in Fig. 4b. Fig. 4c shows the ''dirty'' power spectrum without any RFI excision on a logarithmic scale with the level of RFI about 30 dB above the system noise. Fig. 4d is the waveform of the input signal with a strong RFI source on the same scale as the system noise. Finally, Fig. 4e shows the ''clean'' power spectrum, where the ''spectral line'' is clearly visible and the RFI is significantly suppressed, and Fig. 4g shows the "residuals" or signal-of-interest distortions, that is the difference between "noRFI" power spectrum a) and "clean" power spectrum e). The {\it rms} level of these variations is approximately at the level of system noise {\it rms} $\approx{\sigma^{2}_{sys}}/\sqrt{M}$ in the absence of RFI.
Fig. 5 gives the spectra of four statistical parameters: mean and variance in logarithmic scale, and skew ness and excess in linear scale. Fig. 6 and 7 represent similar computer simulation, but with the RFI not coinciding with the ''spectral line''. Fig. 7c and d show the remarkable characteristics of the skewness and the excess: they are ''transparent'' for the gaussian ''spectral line'' component and react only to RFI presence, that is in the absence of RFI, the skewness = $2$, and the excess = $6$, and for strong RFI both skewness and excess $\to0$. This is in harmony with (10), (11) and Fig. 1a, Fig. 2b, due to "gaussinization" of  probability distribution function: skewness and excess are equal to zero for gaussian distribution.
\begin{figure}
\resizebox{\hsize}{14cm}{\includegraphics{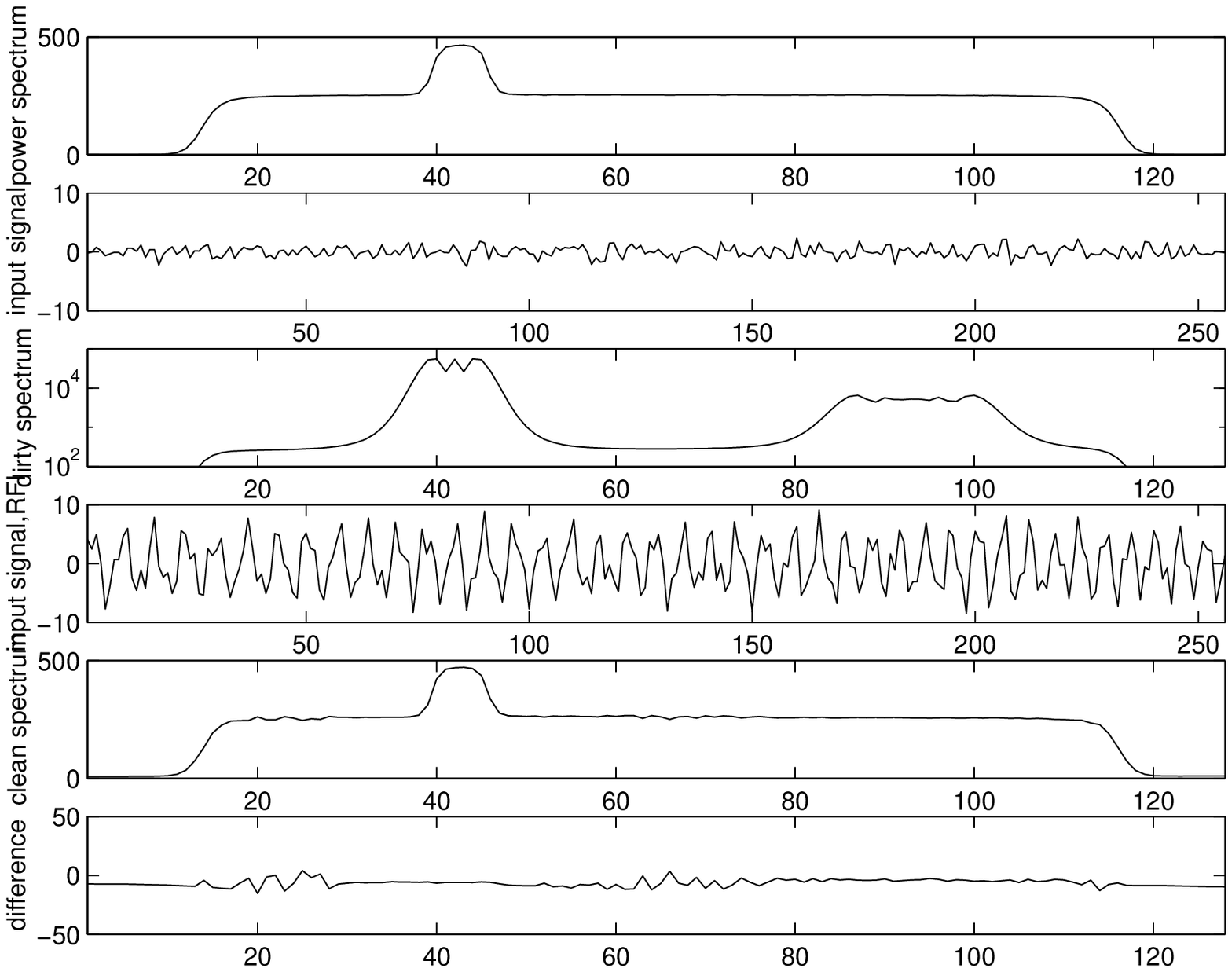}}
\caption{Computer simulations: a) the power spectrum with spectral line with no RFI; b)the input signal with no RFI; c) a two feature RFI power spectrum (logarithmic scale); d)the resulting input signal with RFI; e)the "cleaned" power spectrum; g)the difference between a) and e).}
\end{figure}
\begin{figure}
\resizebox{\hsize}{6cm}{\includegraphics{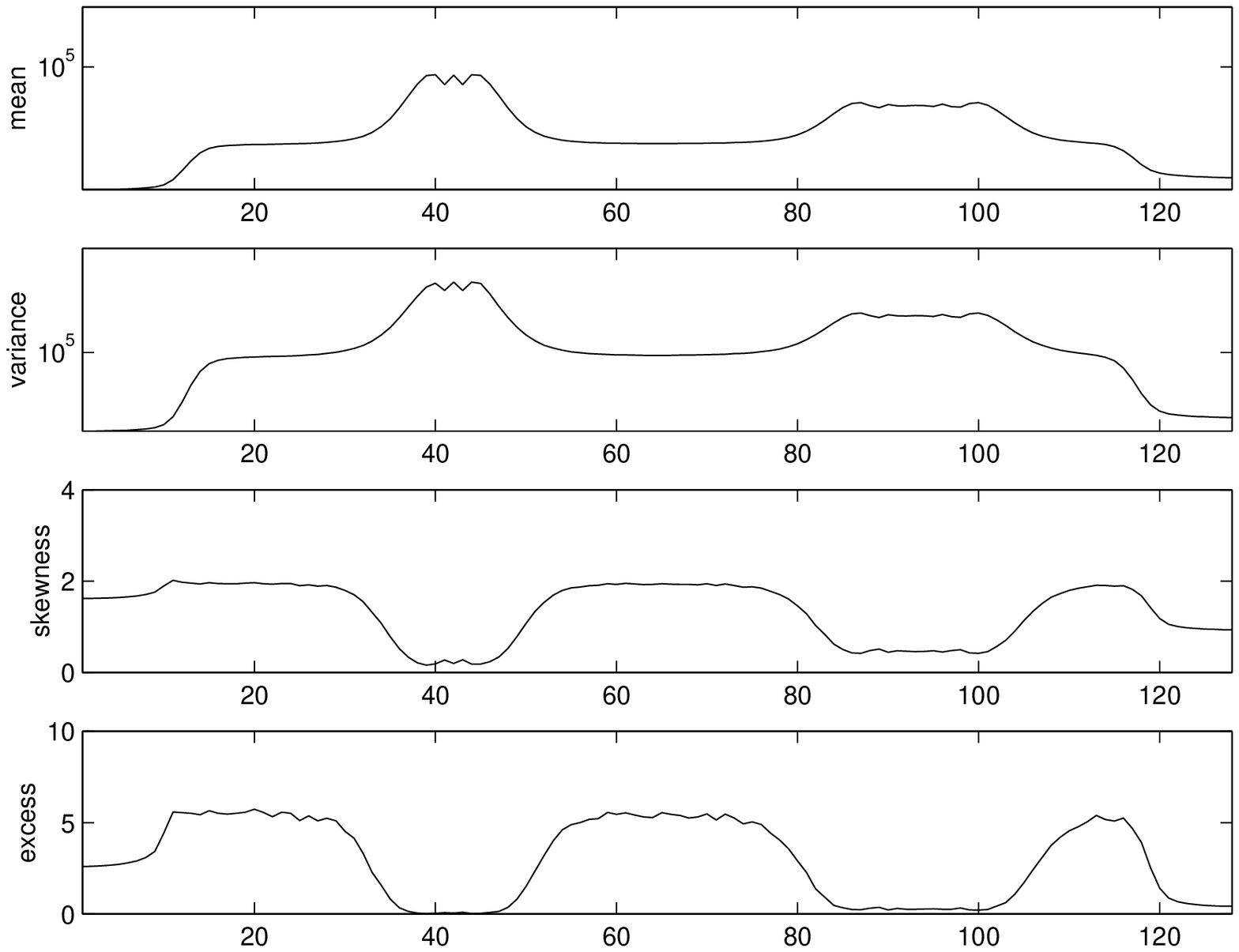}}
\caption{Spectra of the statistical parameters for the model in Fig. 4c (with RFI): a)mean; b)variance; c)skewness; d) excess.}
\end{figure}
\begin{figure}
\resizebox{\hsize}{12cm}{\includegraphics{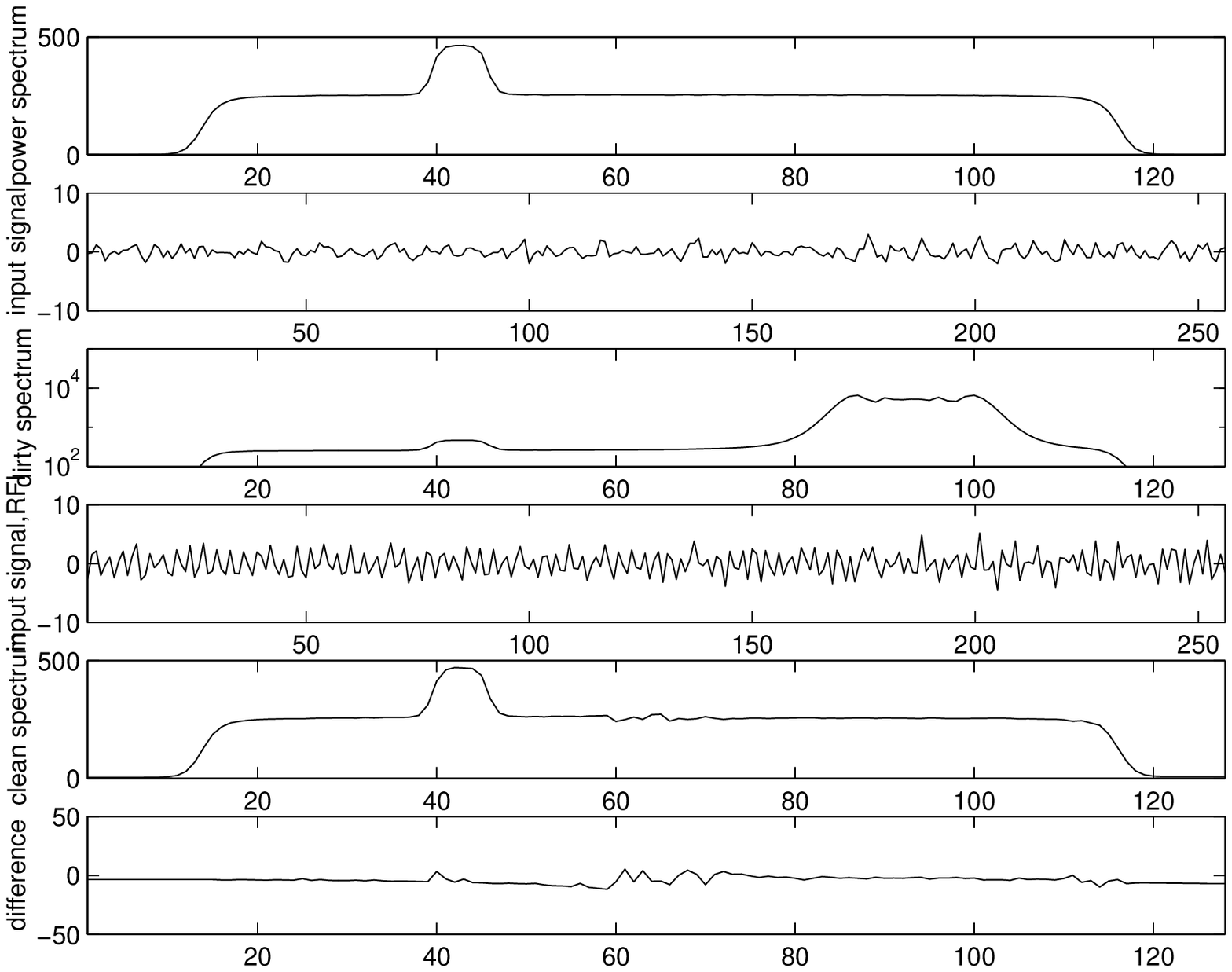}}
\caption{ Computer simulations: a) the power spectrum with spectral line with no RFI; b)the input signal with no RFI; c) a two feature RFI power spectrum (logarithmic scale); d)the resulting input signal with RFI; e)the "cleaned" power spectrum; g)the difference between a) and e).}
\end{figure}
\begin{figure}
\resizebox{\hsize}{6cm}{\includegraphics{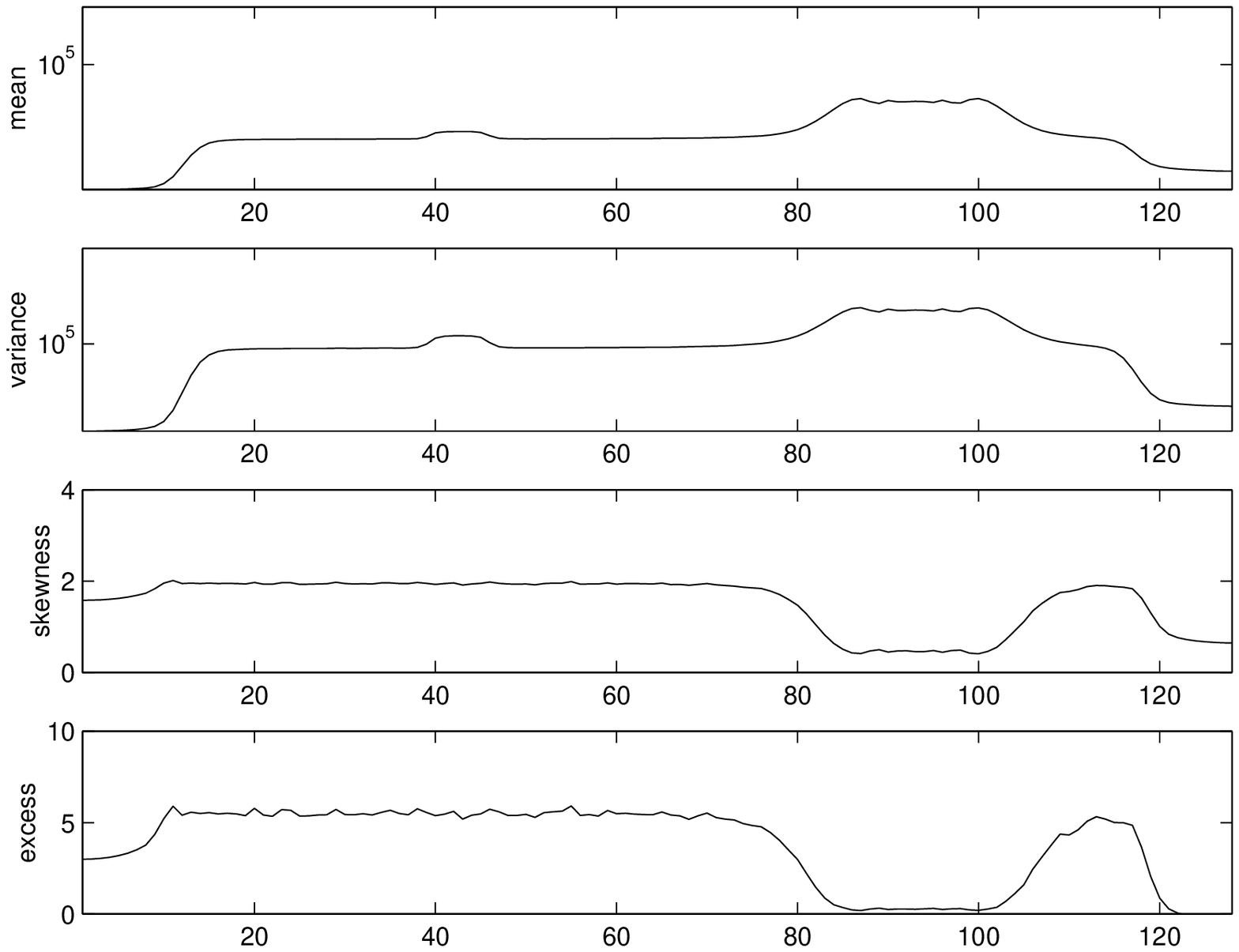}}
\caption{Spectra of the statistical parameters for the model in Fig. 6c (with RFI): a)mean; b)variance; c)skewness; d) excess.}
\end{figure}

\section{Test observations}
Test observations were made with the WSRT to demonstrate the effectiveness of the proposed method with real radio astronomy signals. Digital spectral analysis was performed with the help of a SIGNATEC PMP-8 board bearing eight high performance Texas Instruments DSP TMS320C6201 processors (Fridman \cite{fridman00}). The baseband (0-1.25 MHz) signal at the analogue output of WSRT frequency conversion subsystem corresponding to one of the radio telescopes (RT6) is digitized in a 12-bit analogue-digital converter and then applied to the PMP-8, where the FFT and the averaged statistical parameters were calculated. The final presentation of the ''dirty'' and the ''clean'' spectra, and of their difference were done on a workstation.

Figures 8, 9, and 10 show the results of three test observations of Galactic spectral lines: 21cm HI lines in the direction of Crab and Orion, and an 18cm OH line in the W3(OH) radio source. An artificial continuous wave signal was generated with a Rohde\&Schwarz SMR20 to serve as the RFI signal and was emitted in the direction of RT6 from the WSRT control building. The figures display the data with and without RFI excision applied. These three examples represent relatively strong interfering signals in astronomical data. The RFI suppression obtained in these observational spectra are respectively 17dB in Fig.8, 13dB in Fig.9, and 20dB in Fig.10.
\begin{figure}
\resizebox{\hsize}{10cm}{\includegraphics{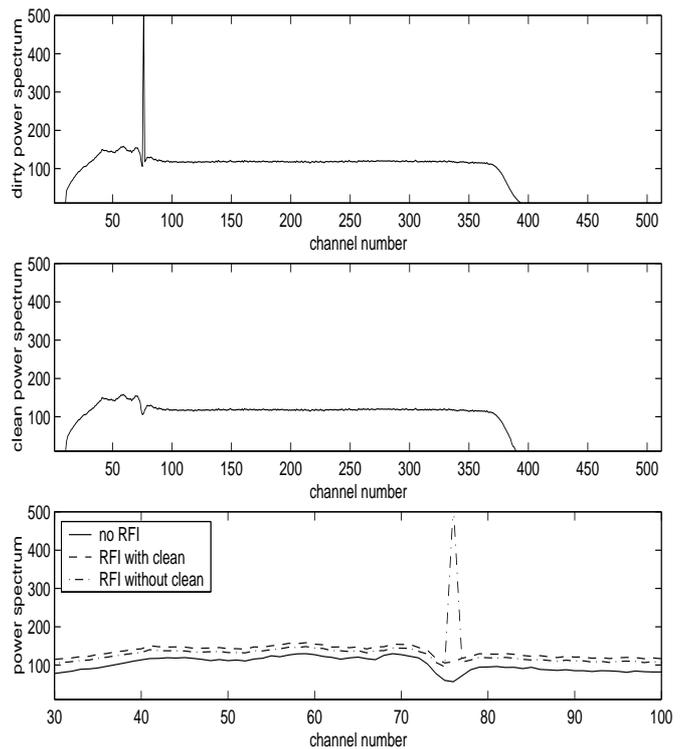}}
\caption{RFI excision using the WSRT with bandwidth = 1.25MHz, sampling frequency = 1.563MHz, and frequency resolution = 3.052kHz: a) a neutral hydrogen spectral line (21cm) in the Galaxy towards the Crab nebula with RFI superposed at 1420.397MHz; b) the same spectrum after RFI excision simultaneously made with frame a; c) a magnification of the spectral line region (channels 30-100) with spectra of a) and b) superposed on a record without any RFI.}
\end{figure}
\begin{figure}
\resizebox{\hsize}{10cm}{\includegraphics{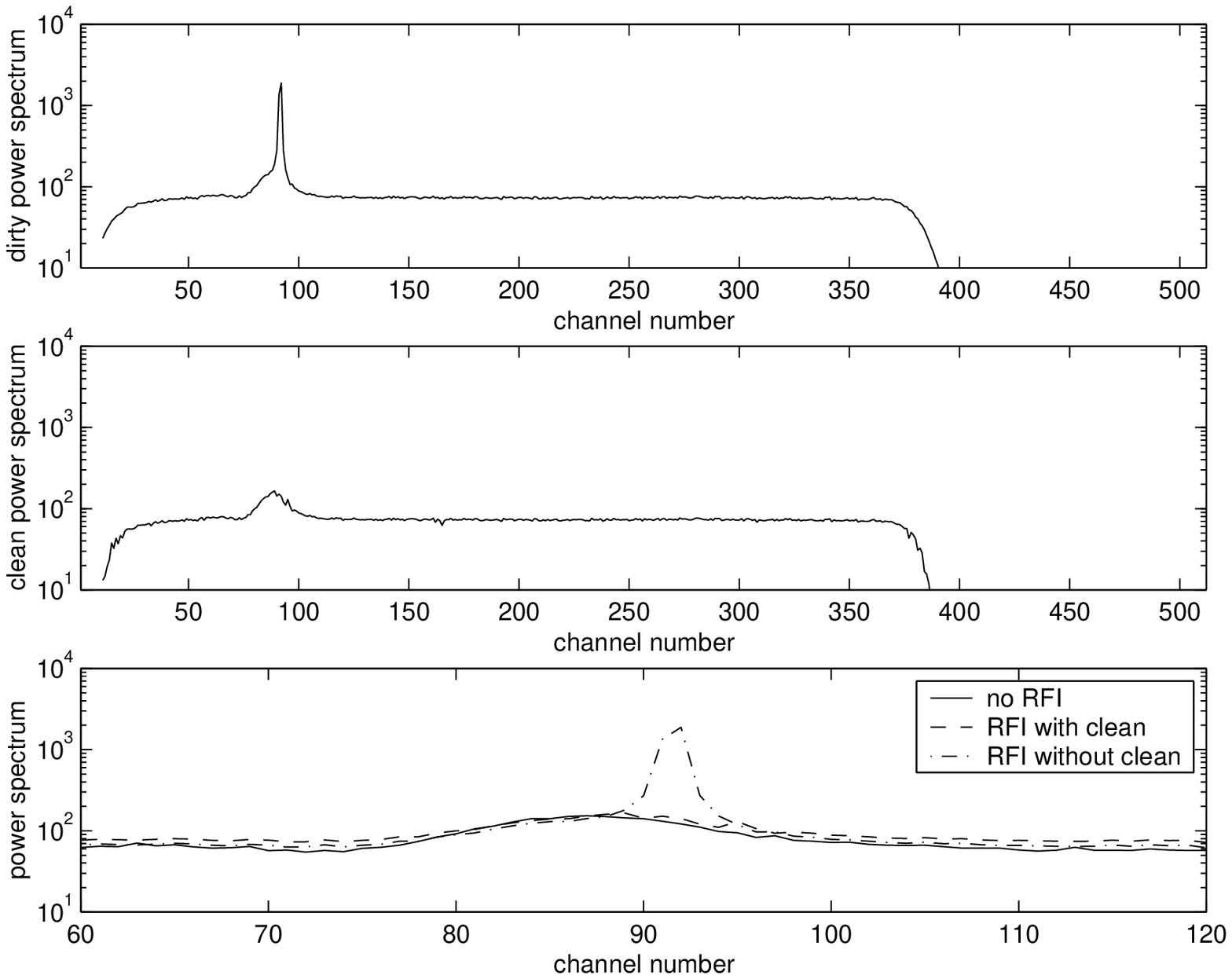}}
\caption{ RFI excision using the WSRT with bandwidth = 1.25MHz, sampling frequency = 1.563MHz, and frequency resolution = 3.052kHz: a) a neutral hydrogen spectral line (21cm) in the Galaxy towards the Orion nebula with RFI superposed at 1420.350MHz; b) the same spectrum after RFI excision simultaneously made with frame a; c) a magnification of the spectral line region (channels 60-120) with spectra of a) and b) superposed on a record without any RFI.}
\end{figure}
\begin{figure}
\resizebox{\hsize}{10cm}{\includegraphics{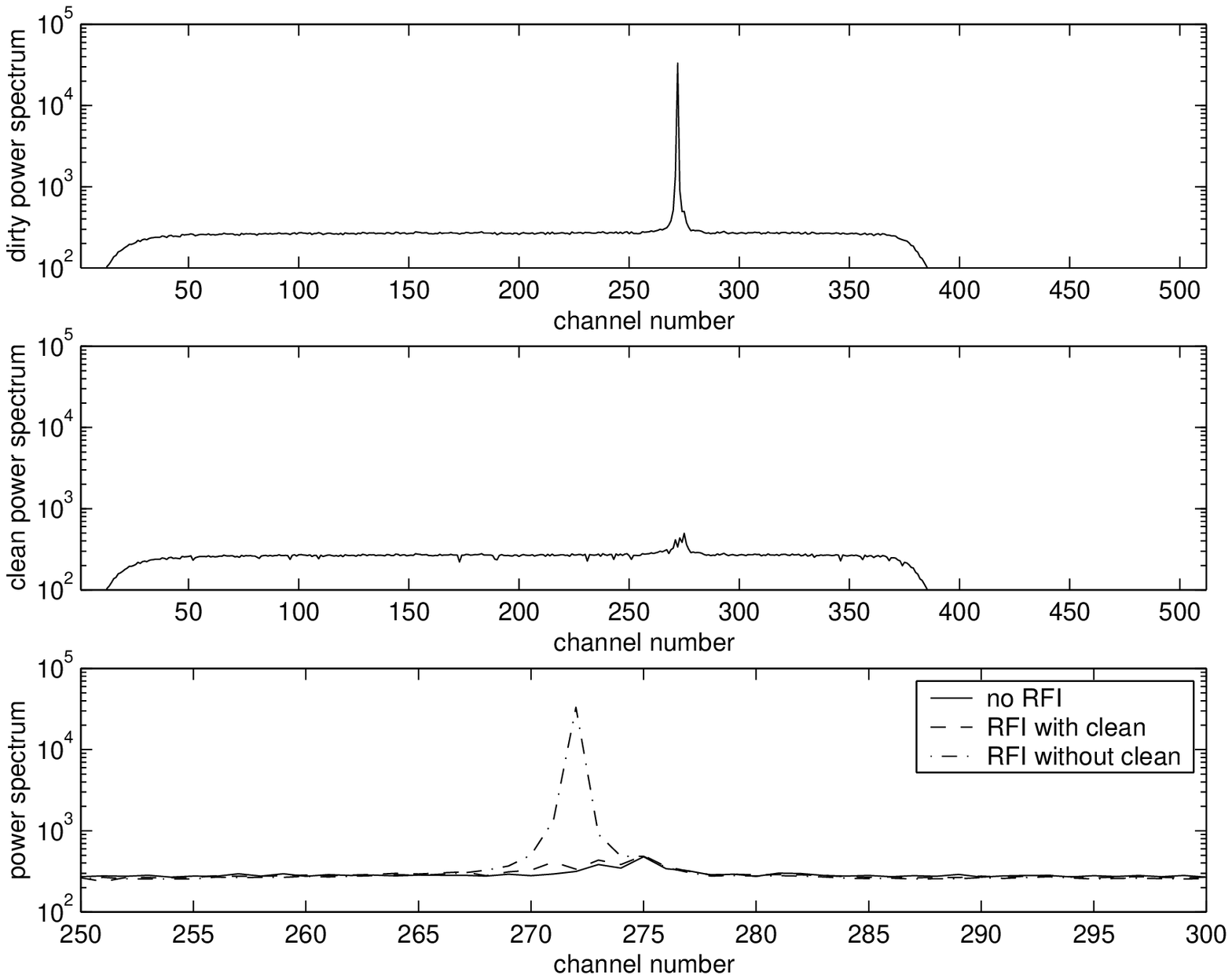}}
\caption{ RFI excision using the WSRT with bandwidth = 2.5MHz, sampling frequency = 3.125MHz, and frequency resolution = 6.104kHz: a) a hydroxyl (OH) emission line (18cm) in the Galaxy towards W3(OH) at 1665.8MHz with RFI superposed; b) the same spectrum after RFI excision simultaneously made with frame a; c) a magnification of the spectral line region (channels 250-300) with spectra of a) and b) superposed on a record without any RFI.}
\end{figure}

\section{Conclusions}
       \begin{enumerate}
\item Single dish spectral observations are most susceptible to RFI. Digital spectral analysis with real-time analysis of the power spectrum higher order statistics provides a promising way for narrow-band RFI excision taking into account the performance of modern DSP systems. The method can be also applied in continuum observations. There is no principal bandwidth limitations, algorithm works in frequency domain, after Fourier transformation. Several RFIs at different frequencies are processed in a parallel way, independently of each other, Fig. 4.
\item  Computer simulations and test observations demonstrate the practical effectiveness of this method with respect to the narrow-band non-gaussian RFI signals.
\item The proposed method is a nonlinear procedure and the post-processing suppression gain (the RFI suppression $INR_{input}/INR_{output}$ ratio)  depends on the input interference/noise ratio $INR_{input} = {A}_{RFI}^{2}/\sigma_{n}^{2}$ in a complex way. For large $INR_{input}$ this gain can be estimated approximately as $10log(INR_{input})+5log(M)$. The stronger RFI, the bigger the gain. The results from our tests with astronomical data show that the gain may range from 13 - 20 dB.
\item The effectiveness of the method depends also on the temporal stability of $INR$: equations (8)-(11) must be solved under the assumption of constant $A$ during the averaging. This limits the value for $M$ and hence the averaging procedure should be divided on three stages: preliminary averaging - RFI subtraction - final averaging.
\item The proposed  method was tested for the  {\it autospectral}, periodogram  analysis, that is  the power spectrum and its high-order statistics were calculated in real time after Fourier transform of the baseband signals. Most of the modern radio astronomy spectrum analyzers use auto- or cross-correlation techniques with 1-2 bit quantization and calculate only the first moment of power spectrum. I think that the direct periodogram spectral analysis with 12-16 bit sampling is more robust to strong RFI - less harmonic distortion products. This is also more suitable for implementation of other RFI mitigation methods using the same DSP hardware, as it is practiced in the RFI mitigation test system  at WSRT.
   \end{enumerate}

\section{Appendix}
For a product of two gaussian values $z=xy$ with variances $\sigma _{1}^{2},\sigma _{2}^{2}$, zero means and correlation coefficient $\rho $ \ the probability distribution function (PDF) is
\begin{eqnarray}
p(z)=\frac{1}{\pi \sigma _{1}\sigma _{2}\sqrt{(1-\rho ^{2})}}K_{0}[\frac{|z|}{\sigma _{1}\sigma _{2}(1-\rho ^{2})}]\times\nonumber\\
\exp [\frac{\rho z}{\sigma _{1}\sigma _{2}(1-\rho ^{2})}],
\end{eqnarray}
and the corresponding characteristic function is
\begin{equation}
f(\xi )=[1-2i\xi \sqrt{\sigma _{1}\sigma _{2}}+\xi ^{2}\sigma _{1}\sigma _{2}(1-\rho ^{2})]^{-1/2}.
\end{equation}

Fig. 11a shows a 3D presentation of how this PDF changes with rising $\rho$. Fig.11b represents several sections of this dependence. It is clear that for $\rho =1$ this distribution is transformed into the distribution of a squared gaussian value.

For a product of two gaussian values $z = xy$ with non-zero means the formula for the PDF is too complex, but the characteristic function is simple enough to calculate the necessary central moments and hence the skewness and the excess. The central moments are more convenient to calculate using cumulants: $\kappa_{r}=i^{-r}\frac{d^{r}}{d\xi^{r}}lnf(\xi)\vert_{\xi=0}$. For variances equal to $1$, the means $a_{1}$, $a_{2}$, and the correlation coefficient $\rho$, the characteristic function is (Deustch\cite{deutsch62}):
\begin{eqnarray}
f(\xi )=\frac{1}{[1-2i\xi \rho +\xi ^{2}(1-\rho ^{2})]^{1/2}}\times\nonumber\\
\exp \{\frac{2i\xi a_{1}a_{2}-\xi ^{2}[a_{1}^{2}-2\rho a_{1}a_{2}+a_{2}^{2}]}{2[1-2i\xi \rho +\xi ^{2}(1-\rho ^{2})]}.
\end{eqnarray}
For the cross-spectral density at a particular frequency, we have to use $f_{\Sigma ^{2}}(\xi )=$ $f(\xi )^{2}$ due to two degrees of freedom, as for the auto-correlation spectrum. Thus the first two central moments corresponding to $f_{\Sigma ^{2}}(\xi)$ and to variances $\sigma _{1}^{2}$, and $\sigma _{2}^{2}$ (so that $\overline{x}=$ $a_{1}\sigma _{1}$, and $\overline{y}=$ $a_{2}\sigma _{2})$ are:
\begin{equation}
m1_{cor}=2\rho \sigma _{1}\sigma _{2}+2\overline{x}\overline{y},
\end{equation}
\begin{eqnarray}
var_{cor}=2\overline{x}^{2}\sigma _{1}\sigma _{2}+2\overline{y}^{2}\sigma _{1}\sigma _{2}+4\rho \overline{x}\overline{y}\sigma _{1}\sigma _{2}+\nonumber\\
2\sigma _{1}^{2}\sigma _{2}^{2}+2\rho ^{2}\sigma _{1}^{2}\sigma _{2}^{2}.
\end{eqnarray}

For $\sigma _{1} = \sigma _{2} = \sigma$, and $\overline{x} = \overline{y} = A/\sqrt{2}$, we have the system of equations with respect to $\sigma$ and $A$:
\begin{equation}
m1_{cor}=2\rho \sigma ^{2}+A^{2},
\end{equation}
\begin{equation}
\mu _{2cor}=var_{cor}=2A^{2}\sigma ^{2}+2\rho A^{2}\sigma ^{2}+2\sigma ^{4}+2\rho ^{2}\sigma ^{4}
\end{equation}
For $\rho\to 1$, we find:
\begin{eqnarray}
m1_{cor}=A^{2}+2\sigma ^{2},\\
var_{cor}=4A^{2}\sigma ^{2}+4\sigma ^{4},
\end{eqnarray}
which is the same as for the auto-correlation spectrum.
\begin{figure}
\resizebox{\hsize}{8cm}{\includegraphics{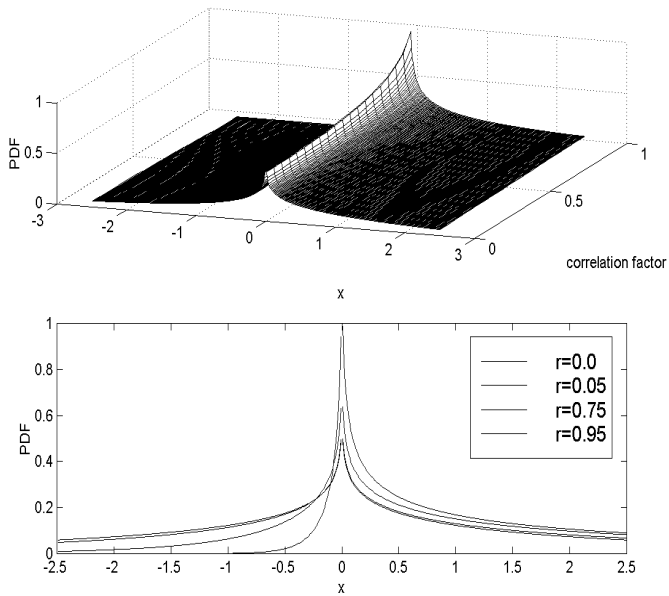}}
\caption{The product probability distribution with the correlation factor as a parameter, a) a 3D presentation; and b) sections for different correlation factors.}
\end{figure}
\begin{figure}
\resizebox{\hsize}{8cm}{\includegraphics{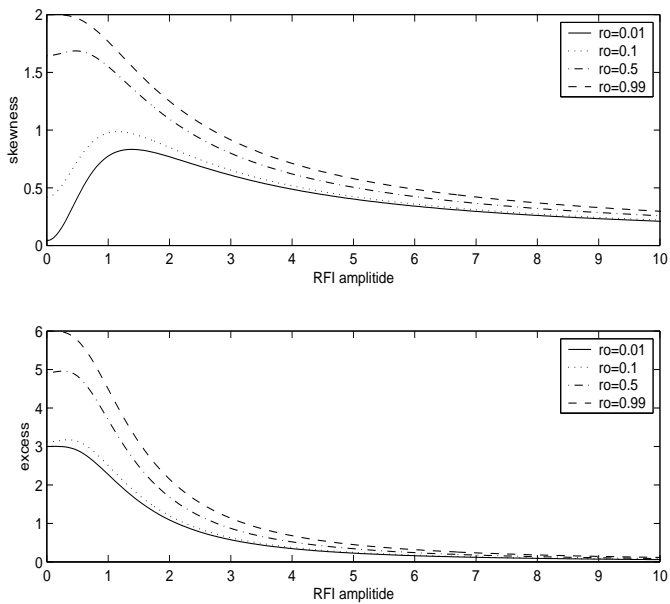}}
\caption{The skewness (a) and the excess (b) for a product with 2 degrees of freedom and $\rho$ as parameter.}
\end{figure}
The following formulae are for the third and the fourth central moments of the sum of two products:
\begin{equation}
\mu _{3cor}=4\rho ^{3}\sigma ^{6}+12\rho \sigma ^{6}+6A^{2}\rho ^{2}\sigma ^{4}+12\rho A^{2}\sigma ^{4}+6A^{2}\sigma ^{4},
\end{equation}
$$\mu _{4cor}=24\rho ^{4}\sigma ^{8}+96\rho ^{2}\sigma ^{8}+24^{2}\sigma ^{8}+48A^{2}\rho ^{3}\sigma ^{6}+96A^{2}\rho ^{2}\sigma ^{6}+$$
\begin{equation}
+96A^{2}\rho \sigma ^{6}+48A^{2}\sigma ^{6}+12A^{4}\rho^{2}\sigma ^{4}+24A^{4}\rho \sigma ^{4}+12A^{4}\sigma ^{4}.
\end{equation}
Using these formulae for $\mu _{i}$, it is possible to calculate skewness
$skew_{cor}=\frac{\mu _{3cor}}{\mu _{2cor}^{3/2}}=$
\begin{equation}
\frac{\sigma ^{4}\sqrt{2}(2\rho ^{3}\sigma ^{2}+6\rho \sigma ^{2}+3A^{2}\rho ^{2}+6\rho A^{2}+3A^{2})}{(\rho ^{2}\sigma ^{2}+\sigma ^{2}+\rho A^{2}+A^{2})^{3/2}},
\end{equation}
and excess
$excess_{cor}=\frac{\mu _{4cor}}{\mu _{2cor}^2}-3=$
\begin{equation}
3\sigma ^{2}\frac{(\rho ^{4}\sigma ^{2}+6\rho ^{2}\sigma ^{2}+\sigma ^{2}+2A^{2}\rho ^{3}+6A^{2}\rho ^{2}+6\rho A^{2}+2A^{2})}{(\rho ^{2}\sigma ^{2}+\sigma ^{2}+\rho A^{2}+A^{2})^{2}}
\end{equation}

When $\rho$ approaches 1, both eqs. (25) and (26) reduce to eqs. (10) and (11), respectively. Fig. 12a,b show how the skewness and the excess depend on the RFI amplitude $A$ with $\rho$ as a variable. When $\rho = 1$ these curves are transformed into the curves in Fig. 2b,c corresponding to the sum of two squares of gaussian values.

\begin{acknowledgements}
The author would like to thank W.A. Baan for helpful suggestions and valuable discussions.
\end{acknowledgements}


\begin{thebibliography}{}

 \bibitem[1996] {ardenne96} Ardenne van, A. and Smits, F.M.A. 1996, in: Proc. of {\it Workshop on Large Antennas in Radio Astronomy}, 28-29 February 1996, ESTEC, Noordwijk, The Netherlands,  59

 \bibitem[1998] {barnbaum98} Barnbaum C., Bradley  R. F. 1998,  The Astronomical J., 115, 2598

 \bibitem[2000] {bregman00} Bregman J.D. 2000, Proc. of SPIE, 4015, 19

 \bibitem[2000] {briggs00} Briggs F. H., Bell J. F., Kesteven M. J. 2000, Astrophysical J., in press.

\bibitem[1999] {cohen99} Cohen, J. 1999, Astronomy\&Geophysics, 40, 6.8

\bibitem[1962]{deutsch62} Deutsch, R. 1962, {\it Nonlinear transformations of random processes}, Prentice-Hall, Inc. Englewood Cliffs, N. J. , Ch. 4

\bibitem[2000]{ellington00} Ellington S. W., Bunton J. D., Bell J. F. 2000, 
 Proc. of SPIE, 4015, 400

\bibitem[1996] {fridman96} Fridman P.A.,  in: Proc. 8th IEEE Signal Processing Workshop on {\it Statistical Signal and Array Processing}, June 24-26, 1996, Corfu, Greece,  264

\bibitem[1997]{fridman97} Fridman P.A. 1997,  Radio Frequency Interference Rejection in Radio Astronomy Receivers, {\it NFRA  Note 664}

 \bibitem[1998]{fridman98}Fridman P.A. 1998, in: Proceedings of {\it EUSIPCO98 Ninth European Signal Processing Conference}, Rhodes, Greece, 8-11 September 1998, 2241

\bibitem[2000] {fridman00} Fridman P.A. 2000, in: Proceedings of the {\it IEEE NORDIC Signal Processing Symposium}, June 13-15,  Kolmarden, Sweden, 375

\bibitem[1999] {kahlmann99} Kahlmann H.C. 1999,  in: Review of Radio Science 1996-1999, ed. by W.Ross Stone, Published for the URSI by Oxford University Press, 751


\bibitem[1997] {lemmon97} Lemmon J.J. 1997, {\it Radio Science}, 32, 525

\bibitem[1999] {leshem99} Leshem, A., van der Veen A. 1999  in: {\it Perspectives on Radio Astronomy Technologies for Large Antenna Arrays}, Smolders, A.B., and van Harlem, M.P., eds., ASTRON, Dwingeloo, 210

\bibitem[1972]{middleton72} Middleton D. 1972, IEEE Trans. on Electromagnetic Compatibility, EMC-14 , 38

\bibitem[1977]{middleton77} Middleton D. 1977,  IEEE Trans.on Electromaghetic Compatibility, EMC-19, 106

\bibitem[1997]{spoelstra97} Spoelstra T. A. Th. 1997,  {\it Tijdschrift van het Nederlands Elektronica- en Radiogenootschap}, 62, 13

\bibitem[1982]{thompson82} Thompson, A. R. 1982,  IEEE Trans. on Ant.\&Propag.,
 AP-30, 450


\bibitem[1971] {whalen71} Whalen A. D. 1971, {\it Detetection of Signals in Noise}, Academic Press, N. Y. , ch. 4

\bibitem[1997] {weber97} Weber, R., Faye, C., Biraud, F., Dansou, J. 1997, 
Astronomy\&Astrophysics Suppl. Ser., 126, 161

\bibitem[1985] {widrow85} Widrow, B. and Stearns, S. 1985, {\it Adaptive Signal Processing}, Prentice-Hall, Inc., Englewood Cliffs, N. J.

\bibitem[1999] {white99} The Elizabeth and Frederick White Conference on
Radio Frequency Interference Mitigation Strategies, December 1999
CSIRO Radiophysics Laboratory Sydney Australia, http://www.atnf.csiro.au/SKA/intmit/atnf/conf/

\end{thebibliography}
\end{document}